\documentclass[a4paper,11pt,DIV=12,draft=false]{scrartcl}

\usepackage[ttscale=0.9]{libertine}
\usepackage[T1]{fontenc}
\usepackage[utf8]{inputenc}

\usepackage{graphicx}
\usepackage{caption}
\usepackage{subcaption}
\usepackage{mhchem}
\usepackage{url}
\usepackage{mathrsfs}
\usepackage{amssymb}
\usepackage{amsmath}
\usepackage{scrlayer-scrpage}
\usepackage[affil-it]{authblk}
\usepackage[numbers,sort&compress]{natbib}
\usepackage[titletoc,title]{appendix}
\usepackage[protrusion=true,expansion,kerning=true,tracking=true,final]{microtype}
\usepackage[
            colorlinks=true,
            linkcolor=hblue,
            citecolor=hgreen,
            filecolor=hblue,
            urlcolor=hred
            ]{hyperref}
\usepackage{enumitem}

\usepackage[dvipsnames,x11names]{xcolor}
\definecolor{hgreen}{rgb}{0,.3,0}
\definecolor{hred}{rgb}{.3,0,0}
\definecolor{orange}{rgb}{1,0.5,0}
\definecolor{hblue}{rgb}{0,0,.3}
\definecolor{LightGray}{gray}{0.95}
\definecolor{gray}{gray}{0.6}

\allowdisplaybreaks
\setcapindent{1em}
\setkomafont{captionlabel}{\bfseries}
\setkomafont{caption}{\itshape}
\setkomafont{titlehead}{\normalsize\sffamily}
\addtokomafont{title}{\boldmath}
\addtokomafont{section}{\boldmath}
\KOMAoptions{headinclude=false,
             footinclude=false,
             parskip=false,
             draft=false,
             overfullrule=false,
             abstract=true,
             numbers=noenddot,
             titlepage=false,
             twocolumn=false,
             twoside=false,
             DIV=12}

\usepackage[shortcuts]{extdash}

\makeatletter
\g@addto@macro\bfseries\boldmath
\makeatother

\makeatletter
\DeclareOldFontCommand{\rm}{\normalfont\rmfamily}{\mathrm}
\DeclareOldFontCommand{\sf}{\normalfont\sffamily}{\mathsf}
\DeclareOldFontCommand{\tt}{\normalfont\ttfamily}{\mathtt}
\DeclareOldFontCommand{\bf}{\normalfont\bfseries}{\mathbf}
\DeclareOldFontCommand{\it}{\normalfont\itshape}{\mathit}
\DeclareOldFontCommand{\sl}{\normalfont\slshape}{\@nomath\sl}
\DeclareOldFontCommand{\sc}{\normalfont\scshape}{\@nomath\sc}
\makeatother

\def\tol#1#2#3{\hbox{\rule{0pt}{15pt}${#1}^{+{#2}}_{-{#3}}$}}

\newcommand{\TeV}{\ensuremath{\text{Te\kern -0.1em V}}}
\newcommand{\GeV}{\ensuremath{\text{Ge\kern -0.1em V}}}
\newcommand{\MeV}{\ensuremath{\text{Me\kern -0.1em V}}}

\newcommand{\Lag}{\mathscr{L}}
\newcommand{\BR}{\mathcal{B}}

\begin{document}
\titlehead{\hfill DO-TH 23/04}
\title{Leveraging on-shell interference to search for FCNCs of the top quark and the Z boson}

\author[a]{Lucas~Cremer%
\thanks{\texttt{lucas.cremer@tu-dortmund.de}}}

\author[b]{Johannes~Erdmann%
\thanks{\texttt{johannes.erdmann@physik.rwth-aachen.de}}}

\author[c]{Roni~Harnik%
\thanks{\texttt{roni@fnal.gov}}}

\author[b]{Jan~Lukas~Sp{\"a}h%
\thanks{\texttt{janlukas.spaeh@physik.rwth-aachen.de}}}

\author[a]{Emmanuel~Stamou%
\thanks{\texttt{emmanuel.stamou@tu-dortmund.de}}}

\date{\today}

\affil[a]{{\large Fakult\"at f\"ur Physik, TU Dortmund, D-44221 Dortmund, Germany}}
\affil[b]{{\large RWTH Aachen University, III. Physikalisches Institut A, Aachen, Germany}}
\affil[c]{{\large Fermilab Accelerator Laboratory, Batavia, IL, USA}}

\maketitle

\begin{abstract}
\normalsize
Flavour-changing-neutral currents (FCNCs) involving the top quark are highly
suppressed within the Standard Model (SM). Hence, any signal in current or
planned future collider experiments would constitute a clear manifestation of
physics beyond the SM.  We propose a novel, interference-based strategy to
search for top-quark FCNCs involving the $Z$ boson that has the potential to
complement traditional search strategies due to a more favourable luminosity
scaling. The strategy leverages on-shell interference between the FCNC and SM
decay of the top quark into hadronic final states.  We estimate the feasibility
of the most promising case of anomalous $tZc$ couplings using Monte Carlo
simulations and a simplified detector simulation.  We consider the main
background processes and discriminate the signal from the background with a
deep neural network that is parametrised in the value of the anomalous $tZc$
coupling.  We present sensitivity projections for the HL-LHC and the FCC-hh.
We find an expected $95\%$ CL upper limit of 
$\mathcal{B}_{\mathrm{excl}}(t\rightarrow Zc) = 6.4 \times 10^{-5}$ 
for the HL-LHC.  In general, we conclude
that the interference-based approach has the potential to provide both
competitive and complementary constraints to traditional multi-lepton searches
and other strategies that have been proposed to search for $tZc$ FCNCs.
\end{abstract}

\newpage
\section{Introduction\label{sec:introduction}}

A flavour-changing-neutral-current (FCNC) process is one in which a fermion changes
its flavour without changing its gauge quantum numbers.
In the Standard Model (SM), FCNCs are absent at tree level, suppressed by
Cabibbo-Kobayashi-Maskawa (CKM) elements, and potentially additionally suppressed
by fermion mass-differences at loop level via the Glashow-Iliopoulos-Maiani
(GIM) mechanism~\cite{Glashow:1970gm}.
The SM predictions for FCNCs that involve the top quark are extremely
small due to the highly effective GIM suppression.
The resulting branching ratios ($\mathcal{B}$) for the top-quark two-body decays via FCNCs range from
$\mathcal{B}(t \rightarrow uH)_{\text{SM}} \sim 10^{-17}$ to
$\mathcal{B}(t \rightarrow cg)_{\text{SM}} \sim 10^{-12}$
~\cite{Eilam:1990zc,Mele:1998ag,Aguilar-Saavedra:2004mfd,Zhang:2008yn,Zhang:2013xya,Forslund:2019fwh}.
However, the top quark plays an important role in multiple theories beyond the SM
due to its large coupling to the Higgs, which is relevant for models
addressing the Hierarchy Problem and models for electroweak-scale baryogenesis.
Several of these models predict enhanced top-quark FCNC
couplings~\cite{Aguilar-Saavedra:2004mfd,TopQuarkWorkingGroup:2013hxj,Azatov:2014lha,Hung:2017tts,Yang:2018utw,Altmannshofer:2023bfk},
which we collectively denote here by $g$.
Typically, constraints on $g$ from low-energy and electroweak-precision observables
are mild~\cite{Larios:2004mx,Han:1995pk,Aranda:2009cd,Gong:2013sh,Gorbahn:2014sha,Hesari:2015oya},
motivating direct searches for FCNC top-quark decays ($t\rightarrow qX$ with $q = u$, $c$)
and FCNC single-top-quark production ($pp \rightarrow tqX$ or $qX \rightarrow t$).
While we focus on FCNC interactions with SM bosons in this paper, FCNC interactions
of the top quark with new, scalar bosons have been proposed~\cite{Castro:2022qkg}
and searched for~\cite{ATLAS:2023mcc}.

Using data taken at the LHC, the ATLAS and CMS collaborations have placed the
most stringent upper limits on top-quark FCNC interactions
via a photon~\cite{ATLAS:2022per,CMS:2015kek}, $Z$ boson~\cite{ATLAS:2023qzr,CMS:2017wcz},
Higgs boson~\cite{ATLAS:2022gzn,CMS:2021hug}, and gluon~\cite{ATLAS:2021amo,CMS:2016uzc}.
Even though many searches take advantage of both the FCNC decay and single production
to search for a non-zero $g$, the limits are traditionally presented in terms of
FCNC branching ratios, $\mathcal{B}(t\rightarrow qX)$.
The most stringent limits at 95\% confidence level~(CL)
range from $\mathcal{B}(t\rightarrow u\gamma) < 8.5\times 10^{-6}$~\cite{ATLAS:2022per}
to $\mathcal{B}(t\rightarrow cH) < 7.3\times 10^{-4}$~\cite{CMS:2021hug}. For FCNCs via
the $Z$ boson, the most stringent limits are obtained in a search that uses the decay of
the $Z$ boson to $e^+e^-$ or $\mu^+\mu^-$ in association with a semileptonically decaying
top quark~\cite{ATLAS:2023qzr}.
The resulting 95\% CL upper limits on $g$ translate to
$\mathcal{B}(t\rightarrow uZ) < 6.2$--$6.6\times 10^{-5}$ and
$\mathcal{B}(t\rightarrow cZ) < 1.2$--$1.3\times 10^{-4}$,
depending on the chirality of the coupling.

While the limits in Ref.~\cite{ATLAS:2023qzr} are obtained with
$\mathcal{L_{\mathrm{int}}} = \int\mathcal{L}\,\mathrm{d}t = 139$~fb$^{-1}$
of data at $\sqrt{s} = 13$~TeV, the HL-LHC is expected to provide
approximately $3000$~fb$^{-1}$ at 14~TeV.
Improved sensitivity to top-quark FCNC processes is hence expected at the HL-LHC,
because statistical uncertainties play an important role in these searches.
With systematic uncertainties being subdominant, one may naively expect that
the upper limits on $\mathcal{B}(t\rightarrow qZ)$ scale with the shrinking
statistical uncertainty.\footnote{Actually, an extrapolation of the ATLAS
FCNC $tZq$ search with $\int\mathcal{L}\,\mathrm{d}t = 36.1$~fb$^{-1}$
at $\sqrt{s} = 13$~TeV~\cite{ATLAS:2018zsq} to the HL-LHC~\cite{ATL-PHYS-PUB-2019-001}
showed a smaller improvement than the naive expectation. This highlights the role of
realistic detector simulations and the consideration of systematic uncertainties
in estimating the sensitivity at the HL-LHC experiments.}
Using this extrapolation, the sensitivity is expected to improve roughly by a
factor $\sqrt{\smash[b]{3000 \, \mathrm{fb}^{-1} / 139 \, \mathrm{fb}^{-1}}} \approx 5$
at the HL-LHC.\footnote{In this rough extrapolation, effects from the more
challenging experimental conditions, improvements due to upgrades to the
detectors, and small changes in the cross sections are neglected.}
The reason for this luminosity scaling is that the partial width for the two-body
top-quark FCNC decay and the cross section for FCNC single production are proportional
to $g^2$ due to the lack of interference with SM processes.\footnote{
Changes in the total top-quark width have a negligible
effect given the current experimental upper limits on $g$.}
As a result, the sensitivity to $\mathcal{B}(t\rightarrow qX)$ naively scales as
$1/\sqrt{\mathcal{L_{\mathrm{int}}}}$ and the sensitivity to $g$ as $1/\sqrt[4]{\mathcal{L_{\mathrm{int}}}}$.
Finding instead an observable that scales linearly with $g$ due to interference
with the SM would modify favourably the luminosity scaling.
Such an interference-based approach would hence be very useful
for the search for top-quark FCNCs.
In the present work we propose such a novel approach and
investigate the feasibility of employing it to search for $tZq$.

\begin{figure}[t]
	\centering
	\includegraphics{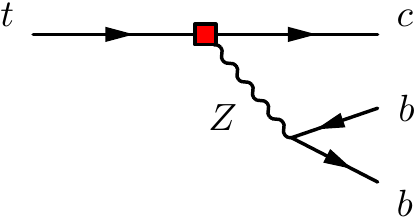}\qquad\qquad
	\includegraphics{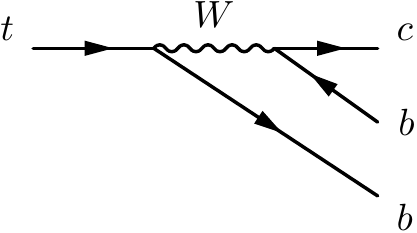}
	\caption{
      The leading-order diagrams for the three-body decay $t\to c b \bar b$.
      The left diagram shows the decay via the FCNC $tZc$ coupling and the right the SM decay via a $W$ boson.
      In the small region of phase space in
      which the $c\bar{b}$-pair reconstructs the $W$-boson mass and the
      $b\bar{b}$-pair reconstructs the $Z$-boson mass, both the $W$ and the $Z$ bosons are
      on-shell and the two amplitudes interfere.
	\label{fig:3bodydecay}
  }
\end{figure}

There are multiple, phenomenologically relevant examples in which
New-Physics (NP) interference with the SM is instrumental for precision NP searches.
Examples include searching for $H\to c\bar c$ via exclusive Higgs decays,
which makes use of interference with the SM $H\to\gamma\gamma$ amplitude \cite{Bodwin:2013gca},
or searching for NP in high-energy diboson distributions by
exploiting the interference between the SM and energy-enhanced NP contributions
from dimension-six operators \cite{Farina:2016rws,Bishara:2020pfx}.
Here, we introduce a new setup that can be applied to improve top-quark FCNCs searches.
As opposed to other approaches, here both NP and SM amplitudes will be mostly resonant,
i.e., contain on-shell --but different-- intermediate particles.
At tree level, a resonant signal amplitude does not generally interfere with a
continuum amplitude, because the former is imaginary and the latter is real.
However, if both the signal and the background contain an on-shell particle,
interference may occur, as long as the final state is identical.\footnote{
For an example of this at the optics table, see Ref.~\cite{Young1804}.} 
In this case of on-shell interference,
NP and SM amplitudes will still interfere, yet the interference will only be large
in a restricted phase-space region.
This potential caveat is different to the ones in the aforementioned examples:
exclusive decays of the Higgs boson are suppressed by the hadronisation probability to
the relevant final-state, e.g., $J/\psi$, and the interference in diboson tails
is suppressed with the decreasing SM amplitude.
Our proposal is to search for the three-body decay $t\rightarrow qb\bar{b}$
in the phase-space region in which there is potentially large NP--SM interference.

The decay $t\rightarrow qb\bar b$ contains two interfering contributions:
the NP contribution $t\rightarrow qZ\rightarrow qb\bar{b}$ and
the SM one $t\rightarrow bW^+\rightarrow qb\bar{b}$, as illustrated in figure~\ref{fig:3bodydecay}.
Consequently, the partial width contains a part that is proportional to $g$.
For sufficiently small $g$ the interference term dominates over the NP$^2$ term ($\propto g^2$) in
which case the sensitivity to $g$ is expected to scale
like $1/\sqrt{\mathcal{L_{\mathrm{int}}}}$, i.e., it improves faster with increasing
luminosity than the traditional approach without interference.
The interference argument also holds for probing the top-quark FCNCs with the Higgs boson ($tHq$)
or with photons ($tq\gamma$) and gluons ($tqg$).
For the Higgs, the interference is suppressed by the light-quark
masses of the final-state quarks ($m_b$ and $m_q$) due to the different chirality structure of
the SM (vector) and NP (scalar) couplings.
For the photon and gluon FCNCs the SM amplitudes peak at small dijet 
invariant masses with potentially large QCD backgrounds,
which require a dedicated study.
We will thus focus in this work on top-quark FCNCs with the $Z$-boson.
We stress that the interference signal is not only sensitive to the magnitude of the $tZq$
coupling but is also sensitive to its phase. The interference approach is hence inherently
complementary to the traditional FCNC searches and of particular interest in case signs of an
anomalous $tZq$ coupling are observed.
We will also focus on the $tZc$ coupling, because the interference is larger compared to $tZu$
due to the larger CKM matrix element $|V_{cb}|$ compared to $|V_{ub}|$.

In section~\ref{sec:interference}, we establish the theory framework
and discuss how to leverage interference based on
parton-level expressions for the interference-based rate and its kinematic properties.
In section~\ref{sec:samples}, we introduce the Monte Carlo (MC) samples that we
use for the sensitivity estimate and discuss the event selection that is tailored
towards the FCNC signal.
In section~\ref{sec:statmethods}, we briefly introduce
the setup of the statistical analysis and then describe in section~\ref{sec:NN}
the optimization of the parametrised deep neural network (DNN) that we use for
the analysis of the simulated data.
The results are given in section~\ref{sec:results-HL-LHC}
for the HL-LHC and in section~\ref{sec:results-FCC} for the FCC-hh.
We present our conclusions in section~\ref{sec:conclusions}.

\section{$t\rightarrow cZ$ from on-shell interference in $t\to c b\bar b$\label{sec:interference}}

The focus of this section is to study the three-body top-quark decay $t\to cb\bar b$ in
the presence of an anomalous, NP $tZc$ coupling with emphasis on how to take
advantage of NP--SM interference to probe the NP coupling.
The decay rate is affected by
interference between the NP and SM amplitudes, illustrated in the
left and right diagram in figure~\ref{fig:3bodydecay}, respectively.
The results of this section are equally well applicable to
the $t\to ub\bar b$ decay when an anomalous $tZu$ coupling is present.
However, this channel is less promising to provide competitive constraints from
an interference-based analysis since the SM amplitude is highly CKM suppressed.
We, thus, concentrate on the $t\to cb\bar b$ case.

Given the smallness of the bottom and charm-quark masses
with respect to the top-quark mass, the NP--SM interference
is large when the chirality of the NP couplings
is the same as the one of the SM $W$-boson contribution, i.e.,
left-handed vector couplings $\bar t_L\gamma^\mu c_L Z_\mu$.
In contrast, the NP--SM interference is suppressed by the
small $b$- and $c$-quark masses if the NP originates from
right-handed vector or tensor operators. Therefore, we only consider here
the most promising case of anomalous left-handed couplings.
The Standard Model Effective Theory (SMEFT) parametrises these couplings in terms
of two dimension-six operators
\begin{align}
  \label{eq:SMEFT}
  \Lag \supset
  \frac{{C^{(1)}_{\varphi q;pr}}}{\Lambda^2}
       (\varphi^\dagger i\overset{\leftrightarrow}{D}_\mu \varphi ) (\bar q_p\gamma^\mu q_r)
     +\frac{{C^{(3)}_{\varphi q;pr}}}{\Lambda^2} (\varphi^\dagger i\overset{\leftrightarrow}{D}{}^a_\mu \varphi )
     (\bar q_p\gamma^\mu \tau^a q_r)\,.
\end{align}
Here, $\varphi$ is the Higgs doublet, $q_p$ left-handed quark-doublets, and $p,r$ flavour indices
in the conventions of Ref.~\cite{Grzadkowski:2010es}.

In the broken phase, by rotating to the quark-mass eigenstates these
SMEFT operators can lead to anomalous tree-level $tZc$ couplings to the left-handed quarks,
which are the subject of this work.
We parametrise them with the phenomenological Lagrangian
\begin{equation}
\label{eq:LagZtc}
\Lag_{tZc} = \frac{g}{2}  e^{i \phi_{\text{NP}}}  ~\bar t_L \gamma^\mu c_L~Z_\mu + \text{h.c.}\,,
\end{equation}
with the NP parameter $g>0$ and the NP phase $0\leq \phi_{\text{NP}}< 2\pi$.\footnote{In unitarity gauge, only the couplings
  in Eq.~\eqref{eq:LagZtc} enter the computation of $t\to cb\bar b$.
  In $R_\xi$ gauges also the corresponding Goldstone couplings must be included.}
In the up-quark mass basis, the coupling in Eq.~\eqref{eq:LagZtc} is related to the
SMEFT Wilson coefficients via $g  e^{i \phi_{\text{NP}}} = \frac{e}{s_wc_w} \frac{v^2}{\Lambda^2}\bigl(C^{(1)}_{\varphi q;32} - C^{(3)}_{\varphi q;32}\bigr)$,
where $e$ is the electromagnetic coupling, $s_w$ ($c_w$) the sine  (cosine) of the weak mixing angle, and $v\simeq 246$\,GeV
the electroweak vacuum-expectation value.\\

The squared amplitude for the $t\to cb\bar b$ decay contains three terms:
the SM$^2$ term, the NP$^2$ term, and their interference, i.e.,
\begin{equation}
  |{\cal A}|^2 = |{\cal A}_{\text{SM}}|^2
  +\underbrace{|{\cal A}_{\text{NP}}|^2}_{\propto g^2}
  + \underbrace{\quad 2 {\rm Re}({\cal A}^*_{\text{SM}} {\cal A}_{\text{NP}})\,,}_{
  \propto ~g \cos(\phi_{\text{NP}}-\phi_{\text{SM}})~\text{and}~g \sin(\phi_{\text{NP}}-\phi_{\text{SM}})}
  \label{eq:Asquare}
\end{equation}
where the underbraces indicate the dependence on the NP parameters.
The interference term depends linearly on the NP coupling $g$ and also
on the relative, CP-violating phase between NP and SM contribution:
\begin{equation}
  \phi\equiv \phi_{\text{NP}} - \phi_{\text{SM}}\qquad\text{with}\qquad \phi_{\text{SM}}\equiv \arg(V_{tb}^*V_{cb})
\end{equation}
As indicated by Eq.~\eqref{eq:Asquare} and further discussed in the following,
the fully differential rate of $t\to cb\bar b$, is sensitive to the interference
term and thus potentially sensitive to both 
a term that is CP-even in the kinematic variables and proportional to $\cos\phi$ 
as well as a term that is CP-odd and proportional to $\sin\phi$.
The cases $\phi = \{0, \pi\}$ lead to a differential rate of $t\to cb\bar b$ that is 
CP conserving. 
In this case, namely, the SM and NP sources of CP violation are aligned and
the differential rate is insensitive to CP violation.

\begin{figure}[t]
  \begin{subfigure}[t]{0.35\textwidth}
    \includegraphics[]{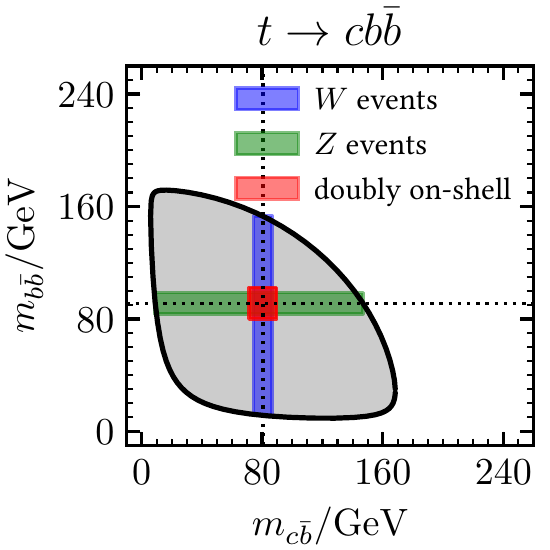}
    \caption{\label{subfig:dalitzfull}}
  \end{subfigure}
  \begin{subfigure}[t]{0.32\textwidth}
    \includegraphics[]{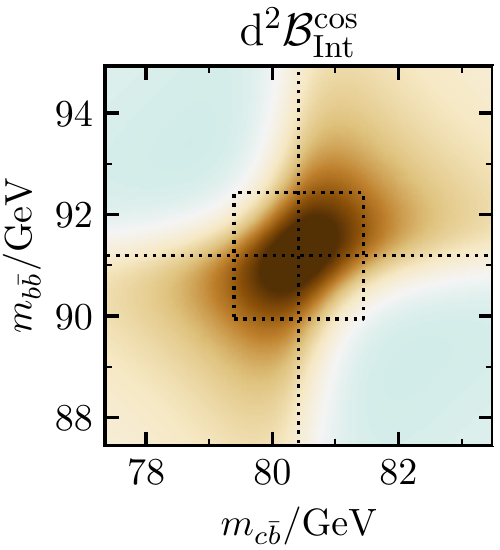}
    \caption{\label{subfig:dalitzcos}}
  \end{subfigure}
  \begin{subfigure}[t]{0.32\textwidth}
    \includegraphics[]{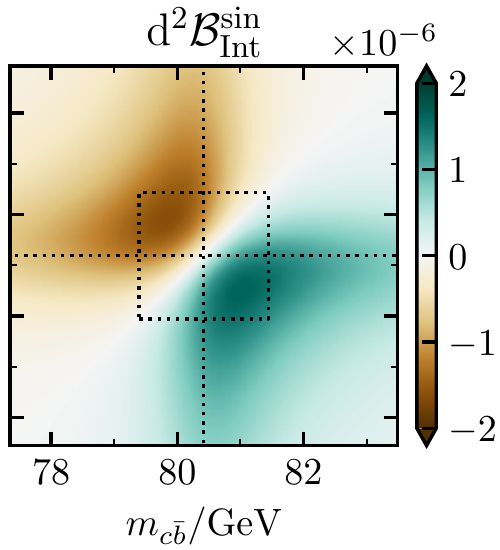}
    \caption{\label{subfig:dalitzsin}}
  \end{subfigure}
  \caption{
    In (\subref{subfig:dalitzfull}) the Dalitz plot for the three-body decay $t\to cb\bar b$ in
    the restframe of the top-quark in terms of the two invariant masses $m_{b\bar b}$ and $m_{c\bar b}$.
    In gray the kinematically physical region. The dotted vertical and horizontal line indicates the
    phase-space points of resonant $Z$- and $W$-boson production (same in (\subref{subfig:dalitzcos}) and
    (\subref{subfig:dalitzsin})).
      ``Pure SM'' events predominantly populate the vertical blue region whereas
      ``pure NP'' events the horizontal green region.
      The red region marks the doubly-on-shell region in which NP--SM interference is
      the largest.
      In (\subref{subfig:dalitzcos}) and (\subref{subfig:dalitzsin}), we show the
      rate originating from NP--SM interference proportional to $g\cos\phi$ and $g \sin\phi$, respectively.
      The figure ranges correspond to the doubly-on-shell region (red region in (\subref{subfig:dalitzfull}))
      and the dotted rectangle centered at the doubly-on-shell point has the width $\Gamma_W$ and the
      height $\Gamma_Z$. Brown regions correspond to negative and green to positive
      contributions to the branching ratio.
    \label{fig:dalitz}}
\end{figure}

The coupling-scaling of the amplitudes does not capture the dependence
on the kinematic variables describing the three-body decay. This dependence
is essential for designing the search that leverages interference in an optimal manner.
The  $t\to cb\bar b$ kinematics are fully specified by the
two invariant masses $m_{c\bar b}^2\equiv(p_c + p_{\bar b})^2$ and
$m_{b\bar b}^2\equiv(p_p + p_{\bar b})^2$.
The different topologies of the NP and the SM amplitudes
(compare the two diagrams in figure~\ref{fig:3bodydecay}) lead to final states with
distinct kinematic configuration:
``SM events'' originate mostly from on-shell $W$'s, i.e.,
$m_{c\bar b} \sim M_W$, whereas ``NP events'' from on-shell $Z$'s, i.e., $m_{b\bar b}\sim M_Z$.
We illustrate this in figure~\ref{fig:dalitz}\subref{subfig:dalitzfull}, which
shows the standard Dalitz plot for the three-body decay in the top-quark rest frame.
in terms of $m_{c\bar b}$ and $m_{b\bar b}$.
The gray area marks the kinematically allowed phase-space.
The SM$^2$ and NP$^2$ parts of the squared amplitude mainly populate the
blue (vertical band) and green (horizontal band) regions, respectively.

The  $W$- and $Z$-boson widths ($\Gamma_W$, $\Gamma_Z$) control
the level of deviations from the on-shell case, i.e., the width of the vertical and
horizontal bands in figure~\ref{fig:dalitz}\subref{subfig:dalitzfull}.
This is best seen by employing the Breit--Wigner approximation for the massive vector
propagators
\begin{equation}
  i\Delta_{\mu\nu}(q) = -i \frac{g_{\mu\nu}-q_\mu q_\nu/M^2}{q^2 - M^2 + i M \Gamma}\,,
  \label{eq:BW}
\end{equation}
which enhances the SM amplitude when $m_{c\bar b} \sim M_W$
and the NP one when $m_{b\bar b}\sim M_Z$.
By integrating over the full phase-space and taking the narrow-width
approximation $\Gamma_W/M_W,\Gamma_Z/M_Z\ll 1$, we recover the usual
relations for the fully inclusive branching ratios originating
from the SM$^2$ and NP$^2$ terms in Eq.~\eqref{eq:Asquare}:
\begin{equation}
  \begin{split}
  |{\cal A}_{\text{SM}}|^2 \propto \BR(t\to cb\bar b)_{\text{SM}} &= \BR(t\to W b)_{\text{SM}}~\BR(W \to c\bar b)_{\text{SM}}\,,\\
  |{\cal A}_{\text{NP}}|^2 \propto \BR(t\to cb\bar b)_{\text{NP}} &= \underbrace{
  \BR(t\to Z c)_{\text{NP}}}_{\propto g^2}~~\BR(Z \to b\bar b)_{\text{SM}}\,,
  \end{split}
  \label{eq:BRincl}
\end{equation}
with $\BR(W \to   c\bar b)_{\text{SM}}\propto M_W/\Gamma_W$
and  $\BR(Z \to   b\bar b)_{\text{SM}}\propto M_Z/\Gamma_Z$.
We collect the expressions for the two-body branching fractions in appendix~\ref{app:BRtwobody}.
However, as we shall demonstrate next, the interference is large in the small phase-space
region in which both $W$ and $Z$ bosons are on-shell (red region in figure~\ref{fig:dalitz}\subref{subfig:dalitzfull}):
\begin{equation}
  M_W - \Gamma_W  \lesssim m_{c\bar b} \lesssim M_W+ \Gamma_W\,,\quad
  M_Z - \Gamma_Z  \lesssim m_{b\bar b} \lesssim M_Z+ \Gamma_Z\,.\quad\text{[doubly-on-shell region]}
\end{equation}
Explicit computation shows that the NP$^2$ and SM$^2$ rates in this
doubly-on-shell region are parametrically suppressed by the widths and masses of the
$Z$/$W$ bosons with respect to their inclusive values in Eq.~\eqref{eq:BRincl}
\begin{equation}
  \BR_{\text{NP/SM}}^{\text{doubly on-shell}} \sim \BR_{\text{NP/SM}}^{}
  \frac{\Gamma_{Z/W}}{M_{Z/W}} \frac{M_{Z/W}^4}{m_t^4}\,.
  \label{eq:BRSMNPdo}
\end{equation}
The net effect is that in total $\BR_{\text{NP/SM}}^{\text{doubly on-shell}}$ are neither
enhanced by $M_{Z/W}/\Gamma_{Z/W}$ nor suppressed by $\Gamma_{Z/W}/M_{Z/W}$ factors,
since $\BR_{\text{NP/SM}}^{}\propto 1/{\Gamma_{Z/W}}$.
The relative suppression, however, is welcome as both of these contributions constitute a
background for the interference-based analysis we are proposing.

In contrast to ``pure SM'' and ``pure NP'' events, ``interference-based'' events
predominantly populate the doubly-on-shell phase-space region, since
$2{\rm Re}({\cal A}^*_{\text{SM}} {\cal A}_{\text{NP}})$ is proportional
to the product of $W$- and $Z$-boson Breit--Wigner propagators.
Summing over final-state polarisations and averaging over the top-quark
polarisation we find the double-differential branching ratio originating
from the interference term in Eq.~\eqref{eq:Asquare} to be
\begin{align}
  \frac{ {\rm d}^2\BR_{\text{Int}}}{ {\rm d} m^2_{b\bar b} {\rm d} m^2_{c\bar b}} &=
  -g \frac{N_{\text{Int}}}{m_t^3 \Gamma _t}
    \frac{\left(m^2_{b\bar b}+m^2_{c\bar b}\right)\left(m_t^2-m^2_{b\bar b}-m^2_{c\bar b}\right)}
  {\bigl((M_W^2-m^2_{c\bar b})^2+\Gamma_W^2 M_W^2\bigr)
   \bigl((M_Z^2-m^2_{b\bar b})^2+\Gamma_Z^2 M_Z^2\bigr)}\biggl[\nonumber\\
   &\qquad\qquad+\cos\phi \left(\left(M_W^2-m^2_{c\bar b}\right) \left(M_Z^2-m^2_{b\bar b}\right) + M_W \Gamma _W M_Z \Gamma _Z\right)
   \label{eq:ddBRint}\\
   &\qquad\qquad+\sin\phi \left(
                  M_Z \Gamma _Z \left(M_W^2-m^2_{c\bar b}\right)
                 -M_W \Gamma _W \left(M_Z^2-m^2_{b\bar b}\right)
                \right)
              \biggr]\nonumber\\
              &\equiv \frac{{\rm d}^2\BR_{\text{Int}}^{\cos}}{ {\rm d} m^2_{b\bar b} {\rm d} m^2_{c\bar b}}\times g\cos\phi
              +  \frac{{\rm d}^2\BR_{\text{Int}}^{\sin}}{ {\rm d} m^2_{b\bar b} {\rm d} m^2_{c\bar b}}\times g \sin\phi\,,\nonumber
\end{align}
with $N_{\text{Int}} = e^3(3-2s_w^2)|V_{cb}||V_{tb}|/(1536\pi^3c_ws_w^3)$.
The last line defines a shorthand notation for the terms proportional to $g\cos\phi$ and $g\sin\phi$.
In figures~\ref{fig:dalitz}\subref{subfig:dalitzcos}
and \ref{fig:dalitz}\subref{subfig:dalitzsin} we show  ${\rm d}^2\BR_{\text{Int}}^{\cos}$ and
${\rm d}^2\BR_{\text{Int}}^{\sin}$, respectively, in terms of the two Dalitz variables.
In brown are the regions with a negative rate and in green the ones with positive rate.
The intersection of the dotted vertical and horizontal line corresponds to the doubly-on-shell point
and we have overlaid a rectangle with width and height equal to $\Gamma_W$ and $\Gamma_Z$.
Eq.~\eqref{eq:ddBRint} and its illustration in figures~\ref{fig:dalitz}\subref{subfig:dalitzcos}
and \ref{fig:dalitz}\subref{subfig:dalitzsin}
contain the most relevant parametric dependences that underpin the idea of leveraging interference
to probe anomalous $tZc$ couplings.
\begin{enumerate}[label=\emph{\roman*)}]
\item\label{it:BW} The denominator in the first line stems from the product of the two Breit--Wigner propagators for the
  $W$ and $Z$ bosons, see Eq.~\eqref{eq:BW}.
  They enhance the rate from interference in the doubly-on-shell region, which is regulated
  by both $\Gamma_W$ and $\Gamma_Z$.
  The enhancement of the doubly-on-shell region with respect to the rest of the phase-space region
  is best seen in figures~\ref{fig:dalitz}\subref{subfig:dalitzcos} and \ref{fig:dalitz}\subref{subfig:dalitzsin}
  for $\mathrm{d}^2\BR_{\text{Int}}^{\cos}$ and $\mathrm{d}^2\BR_{\text{Int}}^{\sin}$.
  The main part of the integrated rate comes from the phase-space region close to
  the doubly-on-shell region.
\item\label{it:phi}
    The rate from interference contains terms proportional to both
    $\cos\phi$ and $\sin\phi$.
    Interference is present independent of whether there is CP violation in
    the decay ($\sin\phi\neq 0$) or whether there is no CP violation ($\cos\phi=\pm1$).
    However, the CP-odd term proportional to $\sin\phi$ is
    odd under the interchange of $W\leftrightarrow Z$ and $m_{b\bar b}\leftrightarrow m_{c\bar b}$
    in Eq.~\eqref{eq:ddBRint}, see also figure~\ref{fig:dalitz}\subref{subfig:dalitzsin} for ${\rm d}^2\BR_{\text{Int}}^{\sin}$.
    The consequence is that the integrated rate
    proportional to $g\sin\phi$ vanishes for the symmetric case $M_W=M_Z$.
    A measurement of the phase $\phi$ thus requires separating
    events within the doubly-on-shell region, which is experimentally extremely challenging
    given the jet energy resolution.
    In contrast, the integrated rate proportional to $g \cos\phi$ is even under the aforementioned
    interchanges and does not vanish after integration,
    see figure~\ref{fig:dalitz}\subref{subfig:dalitzcos} for ${\rm d}^2\BR_{\text{Int}}^{\cos}$.
    A dedicated search in the doubly-on-shell region is thus potentially sensitive to $g\cos\phi$.
\end{enumerate}

In section~\ref{sec:samples}, we will use Monte-Carlo (MC) techniques to simulate events
including a simplified detector simulation
populating the doubly-on-shell region based on the full matrix-elements, which lead
to Eq.~\eqref{eq:ddBRint} and the corresponding expressions for the NP$^2$ and SM$^2$ terms.
To obtain a first rough estimate of the rate from interference and to illustrate
the parametric dependences we present here an approximate phase-space integration
of the rate in Eq.~\eqref{eq:ddBRint}. Most of the rate originates from
events in the doubly-on-shell region, see \ref{it:BW} above. We thus keep
the $m_{b\bar b}$ and $m_{c\bar b}$ dependence in the Breit--Wigner denominators
but set $m_{b\bar b}=M_Z$, $m_{c\bar b}=M_W$ in the remaining squared amplitude.
We then perform the approximate phase-space integration by integrating over
the Breit--Wigner factors via
\begin{equation*}
  \int_{-\infty}^{+\infty}dp^2\frac{1}{(p^2-M^2)^2+M^2\Gamma^2} = \frac{\pi}{\Gamma M}\,,
\end{equation*}
to obtain a rough estimate of the integrated, interference-based rate
\begin{align}
  \label{eq:BRint}
  \BR_{\text{Int}}\approx
  -\pi^2 N_{\text{Int}}\frac{m_t }{\Gamma_t}
    \left(1-\frac{M_W^2}{m_t^2}-\frac{M_Z^2}{m_t^2}\right)
    \left(\frac{M_W^2}{m_t^2}+\frac{M_Z^2}{m_t^2}\right) \times g \cos\phi\,.
\end{align}
We stress that this is only a rough approximation. In fact, the approximation
overestimates the rate by a factor of two with respect to properly integrating
Eq.~\eqref{eq:ddBRint} over the physical kinematic region and including the full
$m_{b\bar b}$ and $m_{c\bar b}$ dependence.

As expected from the discussion in \ref{it:phi} above, Eq.~\ref{eq:BRint} does not contain $g \sin\phi$ terms.
The resulting rate is positive (constructive interference) when
$\cos\phi <0$ and negative when $\cos\phi >0$ (destructive interference), see
colormap of ${\rm d}^2\BR_{\text{Int}}^{\cos}$
in figure~\ref{fig:dalitz}\subref{subfig:dalitzcos}.
For this reason, in the following sections, we will concentrate on
the case of constructive interference by choosing
\begin{equation}
  \label{eq:phipi}
  \cos\phi \equiv \cos(\phi_{\text{NP}} - \phi_{\text{SM}}) \overset{!}{=} -1\,.
\end{equation}
While it may also be possible to search for destructive interference, i.e., a deficit of
events in the doubly-resonant phase space, as for example employed in searches
for heavy scalars \cite{ATLAS:2017snw,CMS:2019pzc} that decay to $t\overline{t}$,
we will not pursue this direction here.
Eq.~\eqref{eq:BRint} also illustrates that $\BR_{\text{Int}}$ is not suppressed by factors of
${\Gamma_{W/Z}}/{M_{W/Z}}$.
As discussed below Eq.~\eqref{eq:BRSMNPdo},
the same holds for the NP$^2$ and SM$^2$ rates in the doubly-on-shell region,
${\BR_{\text{NP/SM}}^{\text{doubly on-shell}}}$.
Therefore, the interference-based rate can compete with
the NP$^2$ rate for sufficiently small $g$ if the analysis targets the doubly-on-shell region.
In what follows we investigate the experimental viability of such a dedicated search.

\section{Simulated samples and event selection\label{sec:samples}}

We generated Monte-Carlo (MC) samples with MadGraph5\_aMC@NLO 3.2.0 (MG5)~\cite{madgraph5}
using a custom UFO~\cite{feynrules} model, which includes the contact $tZc$ coupling
as parametrised in Eq.~\eqref{eq:LagZtc},
setting $\phi=\pi$ (see discussion in Eq.~\eqref{eq:phipi}),
in addition to the full SM Lagrangian with non-diagonal CKM matrix.
All matrix elements are calculated at leading order in perturbative QCD.
We validated the custom model by simulating the decay $t \to c b \bar b$ and comparing the
distribution of events in the two-dimensional plane spanned by the Dalitz variables
$m_{c\bar b}^2$ and $m_{b\bar b}^2$ (cf. section~\ref{sec:interference}) with the
expectation from the explicit calculation (figure~\ref{fig:dalitz}).

In the following, we simulate proton-proton collisions at a centre-of-mass energy of $14\,$TeV.
The structure of the proton is parametrised with the NNPDF2.3LO set of parton distribution functions~\cite{nnpdf23lo1}.
Factorisation and renormalisation scales are set dynamically event-by-event to the transverse
mass of the irreducible $2\to2$ system resulting from a $k_{\mathrm{T}}$ clustering of the
final-state particles~\cite{loop_induced}.
We simulate the FCNC contribution ($\propto g^2$), also referred to as NP$^2$ in Section~\ref{sec:interference}, and the interference contribution ($\propto g$) to
the signal process $t\bar{t} \to c b \bar b\,\mu^-\nu_\mu\bar{b}$ separately,
whereas the SM contribution to this process is treated as irreducible background.
We only simulate the muon channel for simplicity.
The reducible background processes always include top-quark pair
production with subsequent decay in the lepton$+$jets channel with first- or
second-generation quarks $q$ and $q'$.
Besides the six-particle final state ($b\bar q q'\,\mu^-\nu_\mu\bar{b}$),
we also simulate resonant production of additional bottom quarks from $t \bar t Z(\to b \bar b)$
and non-resonant contributions from $t \bar t b \bar b$ and $t \bar t c \bar c$.
We do not simulate several other small background processes, such as $W^{-}+\mathrm{jets}$
production, diboson production with additional jets or $t \bar t H$ production,
because their contribution is expected to be negligible either due to their low cross section or their
very different kinematic properties.

We only generate muons and final-state partons with transverse momenta larger than $20\,$GeV and
require final-state partons to have a minimum angular distance\footnote{
$\Delta R = \sqrt{\left(\Delta\phi\right)^2+\left(\Delta\eta\right)^2}$ with
$\phi$ the azimuthal angle and the $\eta$ the pseudorapidity.}
of $\Delta R = 0.4$ to each other, motivated by the minimum angular
distance obtained with jet clustering algorithms.
We require the same angular distance between final-state partons and
the muon in order to mimic a muon isolation criterion.
For events in the six-particle final state, i.e., signal and background
contributions to $b\bar b b\,\mu^-\nu_\mu\bar{b}$ as well as
the reducible background $b\bar q q'\,\mu^-\nu_\mu\bar{b}$, we require muons and final-state partons
to be in the central region of the detector
($\lvert \eta \rvert < 2.5$).

For simplicity, we do not use a parton shower in our studies.
Instead, we smear the parton-level objects by the detector resolution
in order to approximate detector-level jets, muons, and missing transverse momentum.
The jet resolution is parametrised as $\sigma(p_\text{T})/p_\text{T} = -0.334 \cdot \exp(-0.067 \cdot p_\text{T})+5.788/p_\text{T}+0.039$,
where the transverse momentum, $p_{\text{T}}$, is in units of GeV.
We obtain this parametrisation from a fit to values from the ATLAS
experiment~\cite{atlas_jet_smearing}. We recalculate the energy of each jet
based on the smeared $p_\text{T}$ with the jet direction unchanged.
We smear the $x$- and $y$-components of the missing transverse-momentum vector
independently by adding a random number drawn from a Gaussian
distribution with mean zero and standard deviation of $24$\,GeV~\cite{atlas_met_resolution}.
We then calculate the scalar missing transverse momentum and the corresponding azimuthal angle.
We take the muon transverse momentum resolution to
be 2\% \cite{ATLAS:2016lqx,atlas_muon_momentum_resolution} with no kinematic dependence.

We select events with criteria that are typical for top-quark analyses by the CMS and ATLAS collaborations.
We require the muon to be in the central region of the detector
($\lvert \eta \rvert < 2.5$) and to have a transverse momentum larger
than $25$\,GeV to mimic typical single-muon trigger thresholds~\cite{atlas_muon_trigger_run2,cms_muon_trigger_run2}.
We do not take trigger, identification, or isolation efficiencies into account.
We only accept events with exactly four central jets ($\lvert \eta \rvert < 2.5$)
to reduce the contamination from the reducible background
processes with higher jet multiplicity. Each jet has to have a transverse
momentum larger than $25$\,GeV and we require the missing transverse
momentum to be at least $30$\,GeV.

Given the signal final state, $c b \bar b\,\mu^-\nu_\mu\bar{b}$, we demand the four jets in the event to fulfill the following $b$-tagging criteria.
We require three jets to fulfill a $b$-tagging criterion with a $b$-tagging efficiency of 70\% and corresponding mis-identification efficiencies of 4\% and 0.15\% for $c$-jets and light jets, respectively~\cite{deepjet}.
The additional fourth jet is often a $c$-jet and needs to pass a looser $b$-tagging criterion with a $b$-tagging efficiency of 91\% and a correspondingly larger efficiency for $c$-jets~\cite{deepjet}.
The mis-identification efficiency for light jets of this looser $b$-tagging criterion is 5\%.
We choose the $b$-tagging selection from various combinations of $b$-tagging criteria
with different $b$-tagging efficiencies and corresponding mis-tagging efficiencies.
We choose the combination with the highest value of
$S/\sqrt{S+B}$, where $S$ and $B$ are the total number of weighted events for
the signal and the background contributions, respectively,
as calculated by sampling of jets according to the $b$-tagging efficiencies for the different jet flavours
($S$ contains both the FCNC and interference contribution).

\begin{table}[t]
  \begin{center}
    \caption{The leading-order cross section~$\sigma_{\mathrm{MG}}$ from MG5, the $k$-factors,
             the probability to have only four jets at the LHC for the processes with a six-particle
             final state, $\varepsilon_{\mathrm{4j}}$,
             the fraction of simulated events passing the event selection, $\varepsilon_{\mathrm{pass}}$,
             the $b$-tag efficiency, $\varepsilon_{\mathrm{btag}}$, and the expected number of
             events~$N_{\mathrm{exp}}$ for an integrated luminosity of $\mathit{3000}$\,fb$^{-1}$
             for each process.
             $t \bar t_{\bar{b}c}$ denotes the irreducible SM-background contribution to
             the $b\bar q q'\,\mu^-\nu_\mu\bar{b}$ final state.
             The values of the interference and the FCNC contribution
             are given for $g=0.01$ and $\cos\phi=-1$.\\[-0.5em]
    \label{tab:07_Nexp}
  }
        \begin{tabular}{ccccccc}
        {Process} & {$\sigma_{\mathrm{MG}}$ [pb]} &  {$k$-factor} & $\varepsilon_{\mathrm{4j}}$ & $\varepsilon_{\mathrm{pass}}$ & {$\varepsilon_{\mathrm{btag}}$} & $N_{\mathrm{exp}}$\\
        \hline\\[-1em]
        $\bar t t$                   & $1.73\cdot 10^{1\hphantom{-}}$   & $1.63$ & $0.5$ & $4.4\cdot 10^{-1}$  & $6.7\cdot 10^{-4}$ & $1.24\cdot 10^{4}$\\
        $\bar t t \bar b b$          & $2.29\cdot 10^{-1}$              & $1.17$ & $1$   & $2.8\cdot 10^{-3}$  & $5.7\cdot 10^{-1}$ & $1.27\cdot 10^{3}$\\
        $\bar t t \bar c c$          & $2.12\cdot 10^{-1}$              & $2.41$ & $1$   & $2.8\cdot 10^{-3}$  & $2.9\cdot 10^{-2}$ & $1.21\cdot 10^{2}$\\
        $\bar t t Z$                 & $3.07\cdot 10^{-3}$              & $1.44$ & $1$   & $2.1\cdot 10^{-2}$  & $5.7\cdot 10^{-1}$ & $1.58\cdot 10^{2}$\\
        $\bar t t_{\bar{b}c}$        & $1.46\cdot 10^{-2}$              & $1.63$ & $0.5$ & $4.4\cdot 10^{-1}$  & $1.5\cdot 10^{-1}$ & $2.33\cdot 10^{3}$\\
        Interference                 & $3.35\cdot 10^{-5}$              & $1.63$ & $0.5$ & $4.6\cdot 10^{-1}$  & $1.5\cdot 10^{-1}$ & $5.53\cdot 10^{0}$\\
        FCNC                         & $3.32\cdot 10^{-4}$              & $1.63$ & $0.5$ & $4.6\cdot 10^{-1}$  & $1.5\cdot 10^{-1}$ & $5.58\cdot 10^{1}$\\\hline
\end{tabular}
    \end{center}
\end{table}

Instead of removing events that did not pass the $b$-tagging criteria, we weight events by
the total $b$-tagging probability to avoid large uncertainties due to the
limited size of the MC datasets.
We weight events in samples for the six-particle final states, where we required all four
partons to be central already at generator-level, by a factor of
$\varepsilon_{\mathrm{4j}} = 0.5$, as roughly half of the events in top-quark pair production at the LHC
have more than four jets due to additional radiation~\cite{atlas_ttbar_jet_multiplicity}.
We use $k$-factors to scale the MG5 leading-order cross sections of the MC samples
to higher orders in perturbation theory.
For the six-particle final states associated with top-quark pair production, we use a
value of $986\,\mathrm{pb}$ as calculated at next-to-next-to-leading order in QCD
including next-to-next-to-leading logarithmic soft gluon resummation~\cite{top++}.
For $t \bar t b \bar b$ and $t \bar t c \bar c$, we use cross sections
of $3.39\,\mathrm{pb}$ and $8.9\,\mathrm{pb}$, respectively, as calculated with
MG5 at next-to-leading order~\cite{CMS_ttjets_2020}.
For $t \bar t Z$ production, we use a cross section of $1.015\,\mathrm{pb}$, which
includes next-to-leading order QCD and electroweak corrections~\cite{handbook_4}.
Table~\ref{tab:07_Nexp} summarizes the efficiencies of the event selection,
the MG5 leading-order cross sections, the $k$-factors,
the $b$-tagging efficiencies, and the expected number of events for an
integrated luminosity of $3000\,\mathrm{fb}^{-1}$.

\begin{figure}[t]
	\centering
  \begin{subfigure}[b]{0.48\textwidth}
    \centering
    \includegraphics{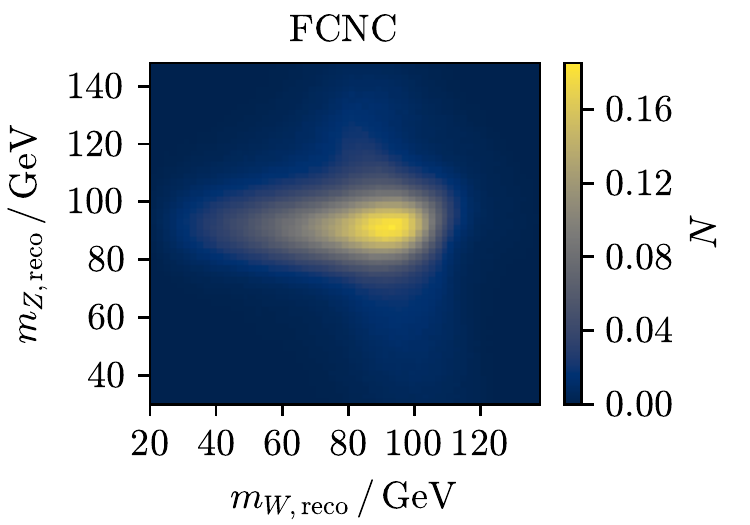}
    \caption{\label{subfig:2d_mass_FCNC}}
  \end{subfigure}
  \hfill
  \begin{subfigure}[b]{0.48\textwidth}
    \centering
    \includegraphics{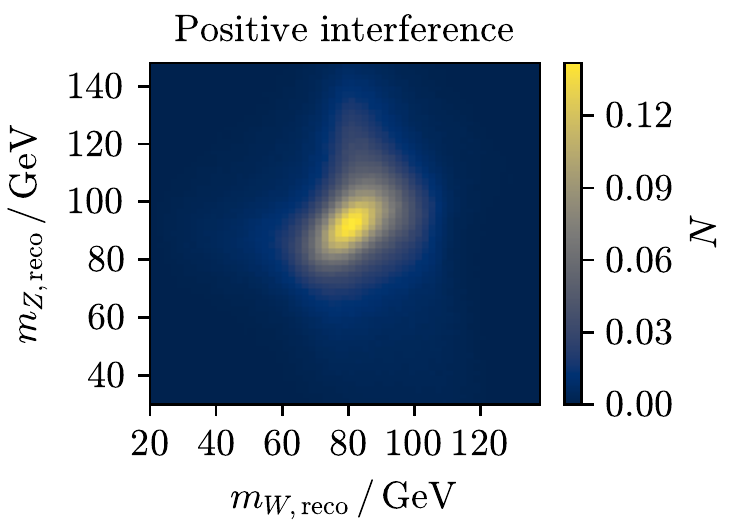}
    \caption{\label{subfig:2d_mass_posInt}}
  \end{subfigure}
  \begin{subfigure}[b]{0.48\textwidth}
    \centering
    \includegraphics{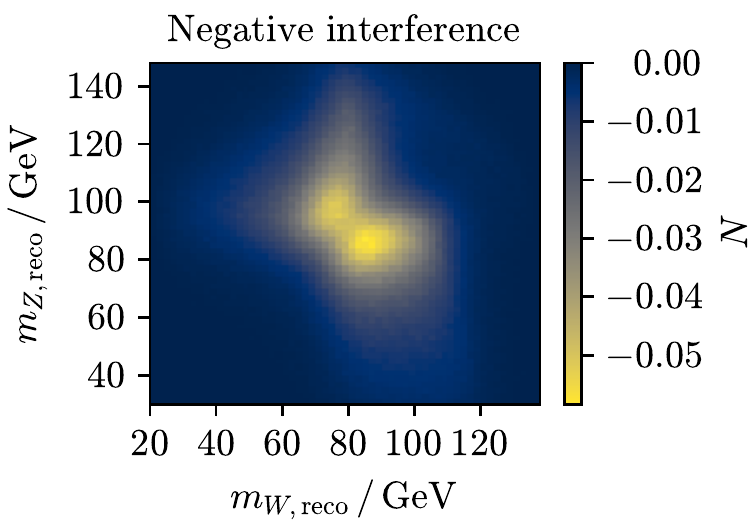}
    \caption{\label{subfig:2d_mass_negInt}}
  \end{subfigure}
  \hfill
  \begin{subfigure}[b]{0.48\textwidth}
    \centering
    \includegraphics{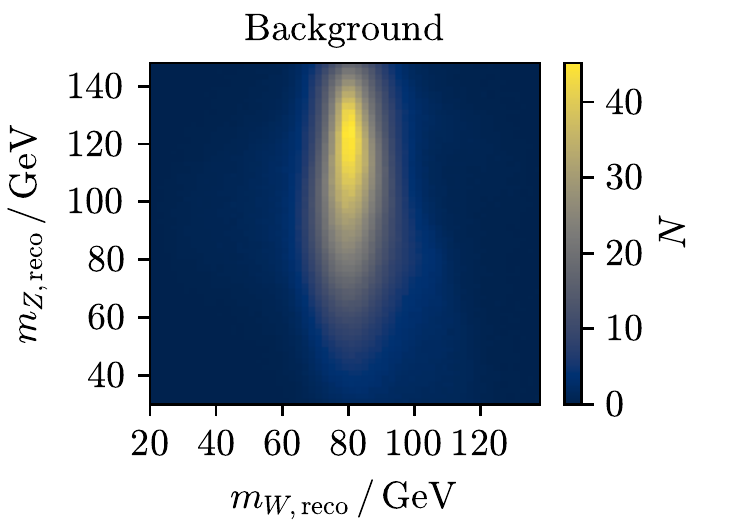}
    \caption{\label{subfig:2d_mass_bkg}}
  \end{subfigure}
  \caption{Expected number of events for $\mathit{3000}$~fb$^{-1}$ in the $m_{W{,}\text{reco}}$
      vs.~$m_{Z{,}\text{reco}}$ plane (in bins of $\mathit{2}$~GeV x $\mathit{2}$~GeV) for the representative value
      $g=0.01$ and $\cos\phi=-1$:
      in (\subref{subfig:2d_mass_FCNC}) from the pure FCNC contribution,
      in (\subref{subfig:2d_mass_posInt}) from the interference contribution with positive
      and in (\subref{subfig:2d_mass_negInt}) with negative event weights,
      and in (\subref{subfig:2d_mass_bkg})~from the sum of the background processes.
    \label{fig:2d_masses}
    }
\end{figure}

To show the detector-level distribution of the expected number of events for $3000\,\mathrm{fb}^{-1}$
we define the variables $m_{W\mathrm{,reco}}$ and $m_{Z\mathrm{,reco}}$
in analogy to  the parton-level Dalitz variables
$m_{c\bar b}$ and $m_{b\bar b}$ (cf. section~\ref{sec:interference}).
For each event, the three jets with invariant mass closest to the top-quark mass
form the hadronically decaying top-quark candidate.
From these three jets, we assume the jet with the lowest sampled $b$-tag score to be the $c$-jet.
In case of a tie, we choose the jet with the higher $p_{\mathrm{T}}$.
The invariant mass of the two remaining jets is $m_{Z{,}\text{reco}}$.
We then calculate the invariant mass of the $c$-tagged jet combined with each of the remaining
two jets of the hadronic top-quark system, and
take the invariant mass closer to $M_W$ as $m_{W\mathrm{,reco}}$.
In figure~\ref{fig:2d_masses}, we show the expected number of events for $3000$~fb$^{-1}$ in the
two-dimensional plane spanned by $m_{W\mathrm{,reco}}$ and $m_{Z\mathrm{,reco}}$
originating from different contributions:
in \ref{fig:2d_masses}\subref{subfig:2d_mass_FCNC} events from the pure FCNC contribution,
in \ref{fig:2d_masses}\subref{subfig:2d_mass_posInt} events from constructive intereference,
in \ref{fig:2d_masses}\subref{subfig:2d_mass_negInt} events from destructive interference,
and in \ref{fig:2d_masses}\subref{subfig:2d_mass_bkg} events from the sum of all
background processes.
The results in figures \ref{fig:2d_masses}\subref{subfig:2d_mass_posInt} and
\ref{fig:2d_masses}\subref{subfig:2d_mass_negInt}
are in qualitative agreement with the parton-level result proportional
to $g\cos\phi$ shown in figure~\ref{fig:dalitz}\subref{subfig:dalitzcos}.
Compared to it, the distributions are more spread out due to the
finite detector resolution.
However, the characteristic differences between pure FCNC, interference,
and background contributions are still visible.

\section{Sensitivity at hadron colliders\label{sec:sensitivity}}

Next, we estimate the sensitivity of the interference-based approach to the $tZc$ FCNC
coupling in the form of expected upper limits on the coupling constant $g$ and
compare it with the traditional approach that focuses on the leptonic decay of the $Z$ boson.
The statistical methodology is briefly outlined in section~\ref{sec:statmethods}.
To separate the FCNC signal, i.e., the pure FCNC contribution, as well as the
interference contribution, from the background, we use a classifier based on
deep neural networks (DNN). We parametrise the DNN as a function of the
FCNC coupling $g$ for optimal separation over a large range of coupling values.
In section \ref{sec:NN}, the architecture and the optimisation of the DNN are explained.
The prospects at the HL-LHC are presented in section~\ref{sec:results-HL-LHC},
and section~\ref{sec:results-FCC} contains estimates for the sensitivity to $g$
in various future scenarios.
The section concludes with a comparison to other approaches to constrain $tZc$ FCNC couplings
in section~\ref{sec:other_approaches}.

\subsection{Outline of the statistical methods\label{sec:statmethods}}

Our metric for the sensitivity to the $tZc$ FCNC coupling is the 95\% CL expected upper limit on $g$ since this allows for a straightforward comparison with existing searches.
The method to derive the upper limit is the following:
We create pseudo-measurements by sampling from the background-only histogram assuming
a Poisson distribution for the counts per bin.
Motivated by the Neyman-Pearson lemma~\cite{neyman_pearson},
we construct a likelihood-ratio test statistic, $t$, by comparing the bin counts from the pseudo-measurements
$\vec{x}$ with the expectation values from the MC simulation under the
$s\!+\!b$-hypothesis ($b$-only-hypothesis) $\vec{\lambda}_{s+b}$ ($\vec{\lambda}_{b}$)
for each pseudo-measurement:
\begin{equation}
  t = - 2 \ln\left( \frac{\mathcal{L}(\vec{x} \mid \vec{\lambda}_{s+b})}{\mathcal{L}(\vec{x} \mid \vec{\lambda}_{b})} \right)\,,
      \quad\text{with}\quad\mathcal{L} = \displaystyle \prod_{i=1}^{N_\text{bins}} \frac{\lambda_i^{x_i}}{x_i!}\mathrm{e}^{-\lambda_i}\,.
\end{equation}
The nominal expected upper limit on the coupling strength, $g_\mathrm{excl}$, is derived as the median
of all pseudo-measurements under the assumption of the absence of a signal with the
CL$_\mathrm{s}$ method~\cite{cls}.

\subsection{Optimisation of the parametrised deep neural networks\label{sec:NN}}

Resolution effects, in particular the jet-energy resolution, and wrong assignments
of jets to the decay branches complicate the reconstruction of invariant
masses at detector level and motivate the use of machine-learning techniques
to optimise the separation of signal and background in a high-dimensional space.
We use the following 31 variables for the training of the DNN: for the $b$-tagged jets,
their transverse momenta, pseudorapidities, azimuthal angles, energies and
the highest-efficiency $b$-tagging working point that the jet passes;
for the single muon, its transverse momentum, pseudorapidity and azimuthal angle;
for the missing transverse momentum, its magnitude and azimuthal angle.
The values of all azimuthal angles $\phi$ are replaced by the combination
of $\sin\phi$ and $\cos\phi$ due to the periodicity of the azimuthal angle.
The natural logarithm is applied to all transverse momentum and energy spectra and the
missing transverse momentum spectrum, as these variables have large positive tails.
The dataset is split with fractions of 60\%~:~20\%~:~20\% into training, validation and test sets.
As a last step, all variables are studentised using $y_i' = (y_i - \mu)/\sigma$,
where $\mu$ refers to the arithmetic mean of the respective
variable and $\sigma$ is the estimated standard deviation.

Besides these 31 observables, we also use the coupling constant $g$
as an input to the DNN, which leads to a parametrised DNN \cite{PNN}.
The idea is to present different values of $g$ to the DNN during the training so that the
DNN learns the relative importance of the different signal contributions as a function of $g$.
For example, for $g \gtrsim \mathcal{O}(0.1)$ the DNN should not focus on the interference
contribution at all and instead concentrate on the separation of the FCNC contribution
against the backgrounds.
This is because the weight of the FCNC contribution exceeds that of the interference
contribution by orders of magnitude in that regime.
Conversely, for $g \lesssim \mathcal{O}(0.001)$ the DNN should start to focus on the
interference contribution more and more to leverage the slower decrease of the number
of expected events for the interference contribution compared to the FCNC contribution.
To give the DNN the possibility to learn this dependence, we further split the
training and the validation set into five stratified subsets. Each of these subsets
corresponds to a specific value of $g\in \{0.001,\,0.005,\,0.01,\,0.05,\,0.1\}$.
These values are chosen to cover the range around the current best exclusion limit
of about $0.0126$~\cite{ATLAS:2023qzr}.
For the training, the weights of the signal events are adjusted so that for a given
value of $g$ the sum of weights in each subset corresponds to the sum of
weights of the background contribution.

The constructed DNN has four output nodes: one for pure FCNC events,
one for interference events with positive weight, one for interference
events with negative weight, and one for background events.
For the output layer, we use {\tt softmax} and for the hidden layers {\tt ReLU} as the activation function.
We use the Adam optimiser~\cite{adam} and categorical cross-entropy as the loss function.
For the determination of the expected exclusion limit, a one-dimensional discriminant
\begin{equation}
    \label{eq:discriminant}
    d = \frac{1-\alpha_{\mathrm{bkg}}-\alpha_{\mathrm{negInt}}+
        \alpha_{\mathrm{posInt}}+\alpha_{\mathrm{FCNC}}}{2} \in [0{,}1]
\end{equation}
is constructed based on the activation~$\alpha$ of the respective output nodes.
We assign a negative prefactor to the output node
corresponding to the negative interference contribution, to increase the difference between the background-only and the signal distribution of $d$.
The corresponding histograms of $d$ consist of 10 equidistant bins.
To account for charge-conjugated processes, the bin contents are multiplied by a factor of two.

The structure of the DNN as well as the learning rate and the batch size during the
training are manually optimised based on the expected exclusion limit on the validation
set.
A learning rate of 0.001 and a batch size of 1000 is chosen.
The final structure of the DNN is $[32,\,128,\,256,\,128,\,64,\,32,\,4]$,
with the numbers referring to the number of nodes in the respective layer. The evolution of the expected exclusion
limit during the training of the DNN are shown in figure~\ref{fig:train_hist_and_CLs}\subref{subfig:train_exp_limits_plot}.

\subsection{Prospects for HL-LHC\label{sec:results-HL-LHC}}

\begin{figure}
  \begin{subfigure}[t]{0.5\textwidth}
    \centering
    \includegraphics{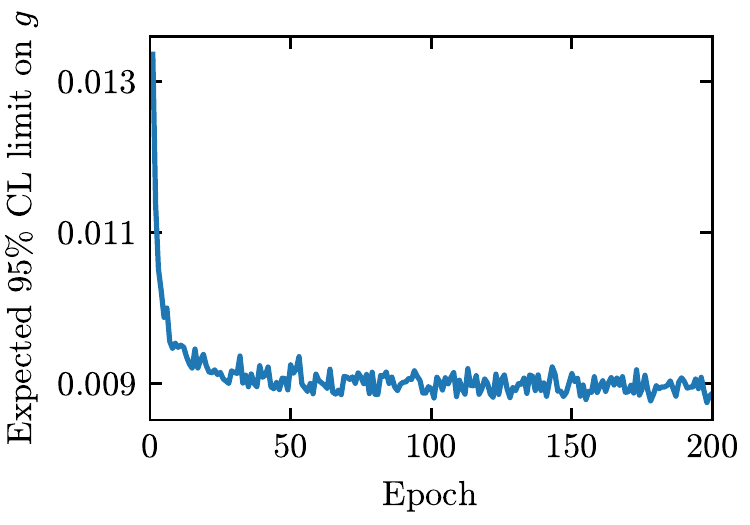}
    \caption{\label{subfig:train_exp_limits_plot}}
  \end{subfigure}\hfill
  \begin{subfigure}[t]{0.5\textwidth}
    \centering
    \includegraphics{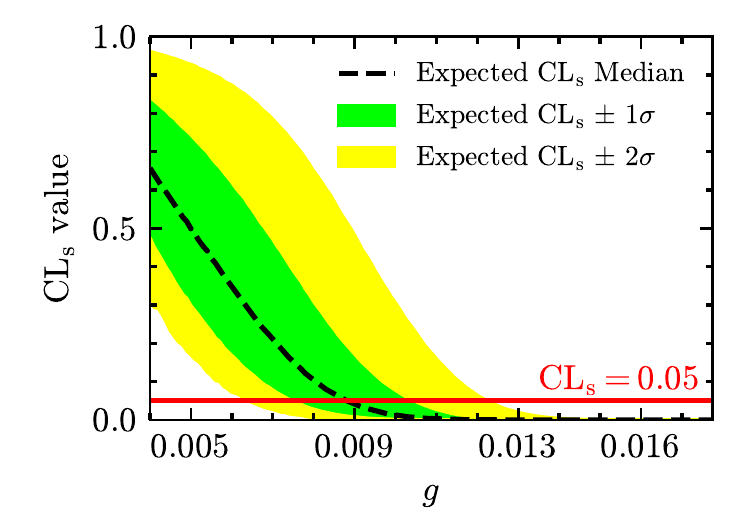}
    \caption{\label{subfig:CLsPlot}}
  \end{subfigure}
  \caption{
    In (\subref{subfig:train_exp_limits_plot}), the expected 95\% CL exclusion limit on $g$ calculated on the
      validation set after each epoch during the training of the DNN.
      In (\subref{subfig:CLsPlot}), the CL$_{\mathrm{s}}$ value estimated for various values of the coupling
    constant~$g$ and the corresponding $\pm 1\sigma$ and $\pm 2\sigma$
    uncertainty bands.
   \label{fig:train_hist_and_CLs}
   }
\end{figure}

The integrated luminosity expected at the HL-LHC is $\mathcal{L}=3000\,\mathrm{fb}^{-1}$~\cite{HLLHC}.
Figure~\ref{fig:train_hist_and_CLs}\subref{subfig:CLsPlot} contains the
CL$_{\mathrm{s}}$ values resulting from the evaluation of the DNN on the test
as a function of the coupling constant $g$.
We find an expected upper exclusion limit at 95\% CL of
\begin{equation}
  g_{\mathrm{excl}}= \tol{8.8}{1.7}{1.3}\times 10^{-3}.
\end{equation}
The corresponding nominal upper limit on the branching fraction
is $\mathcal{B}_{\mathrm{excl}}(t \rightarrow Zc) = 6.4 \times 10^{-5}$.

\begin{figure}[t]
  \centering
  \includegraphics{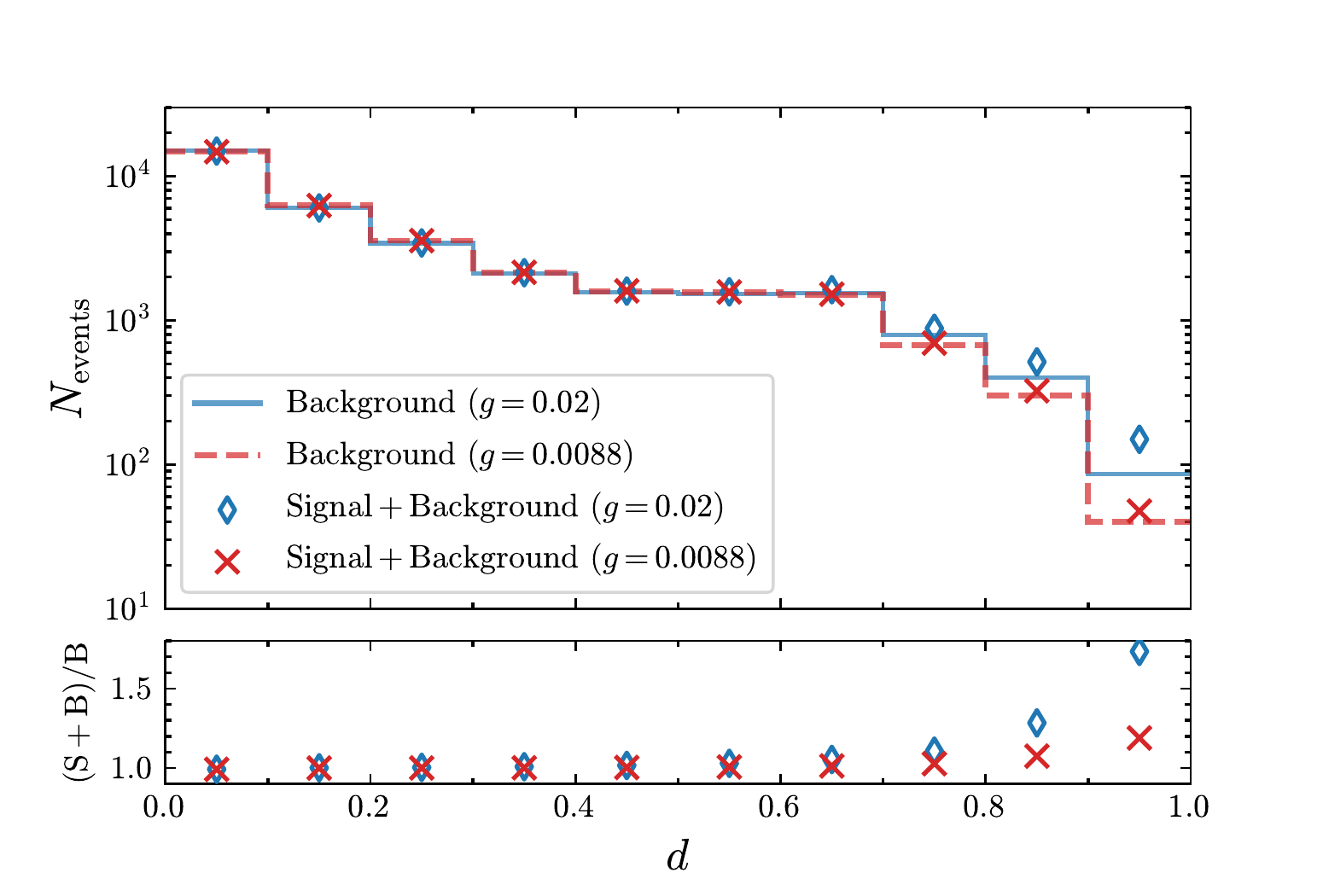}
  \caption{
    The signal and background distribution of the discriminant for $g=8.8\times 10^{-3}$
    and $g=0.02$. As the DNN is parameterised in $g$, the background distribution depends
    on $g$ as well.
    The bottom panel shows the ratio of expected signal+background events divided
    by the number of expected background events, $(S+B)/B$.
    \label{fig:disc_comb}
  }
\end{figure}

In the following, we highlight some of the features of the machine-learning
based analysis to illustrate the employed methods.
The distributions of the discriminant for $g=g_{\mathrm{excl}}$ and the
rejected hypothesis $g=0.02$ are shown in figure~\ref{fig:disc_comb} for the signal
and the background-only hypothesis. Since the DNN is parameterised in $g$, the
background-only distribution depends on $g$ as well. The number of background events
expected in the rightmost bins increases for $g=0.02$ compared to the bin contents
expected for $g=g_{\mathrm{excl}}$. This implies that the DNN adapts to the
simplifying kinematics due to the decreasing importance of interference events.

\begin{figure}[t]
  \centering
  \includegraphics{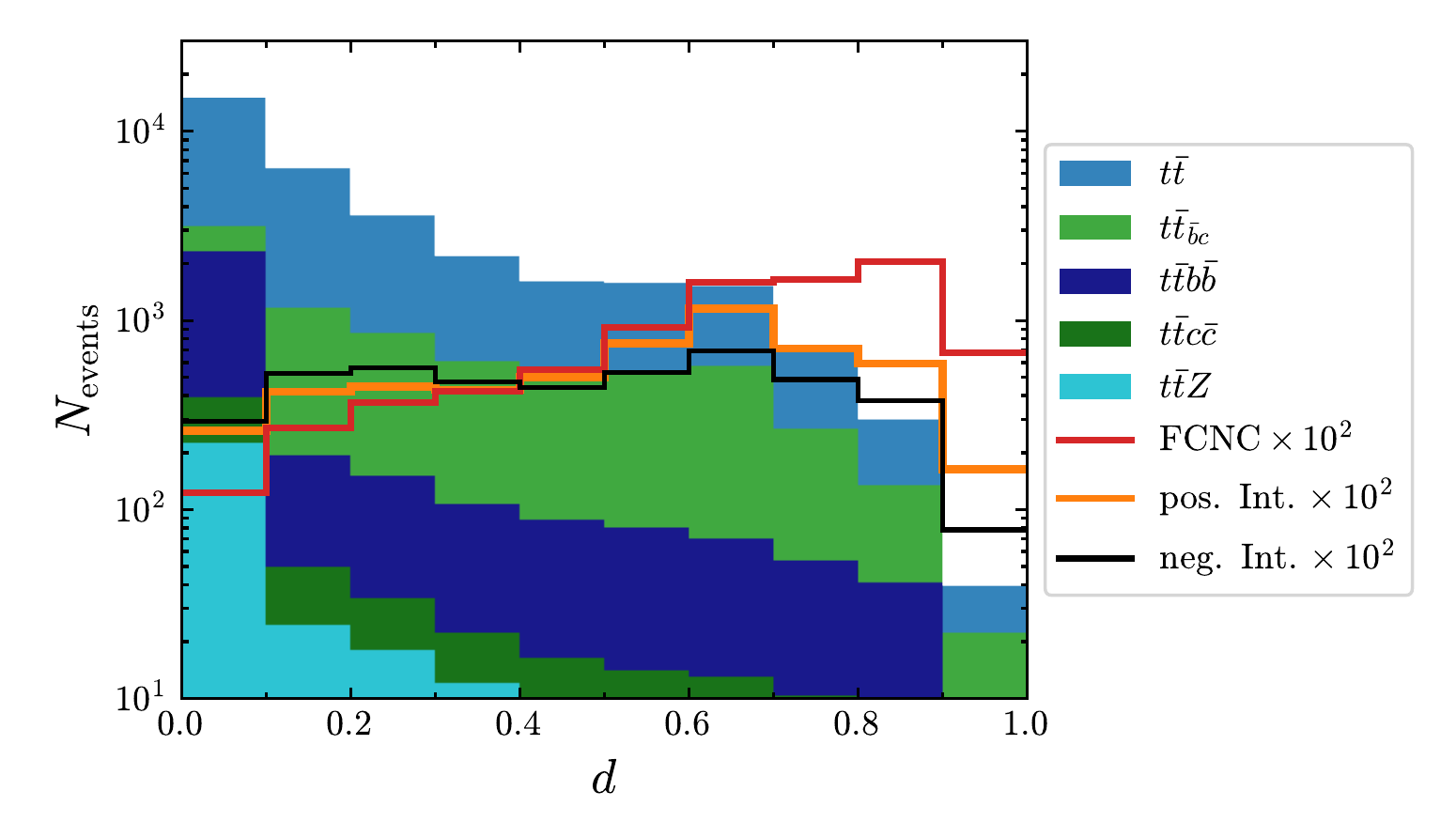}
  \caption{
    Number of events for each background process in bins of the discriminant $d$.
    The expected number of events in each bin is determined from the nominal expected
    exclusion limit $g = 8.8 \times 10^{-3}$ and an integrated luminosity of 
    $\mathit{3000}$\,fb$^{-1}$ at HL-LHC.
    In addition, the shapes of the signal distributions are illustrated.
    \label{fig:disc_split}
  }
\end{figure}

In figure~\ref{fig:disc_split} we show both the bin contents expected
for $g=g_{\mathrm{excl}}$ for each background process and the
shapes of the signal contributions.
Since the irreducible SM background $t\overline{t}_{\overline{b}c}$ has the same
final state as the signal, the separation from signal events turns out to be rather
difficult compared to the reducible backgrounds.
In fact with respect to the aforementioned irreducible component,
the separation of top-quark pair production with decays to only first-
and second-generation quarks, denoted by $t\overline{t}$, can be separated better.
Nevertheless, this process remains the most important background contribution
due to its high cross section.

The DNN separates the signal from the three processes with an additional
heavy-flavour quark pair well; this can be attributed to the different kinematical
structure due to the additional particles in the event.
It should also be noted that the FCNC distribution has a slightly higher mean
than the positive interference distribution.
This is due to two factors:
Firstly, in the vicinity of $g=g_{\mathrm{excl}}$ the sum of weights of the FCNC contribution
is still a bit larger than the sum of weights of the positive interference contribution.
Thus, the DNN focusses on separating the FCNC events from the background events
because of their larger relative impact on the loss function.
Secondly, the distribution of the events in the considered phase space inherently
offers more separation power from the background for the FCNC events compared to
the interference events, as visualised in the $m_{W{,}\text{reco}}$ vs.\ $m_{Z{,}\text{reco}}$
plane shown in figure~\ref{fig:2d_masses}.
Additionally, the mean value of the distribution for negative interference events
is only slightly lower compared to the positive interference contribution, even
though the definition of the discriminant in Eq.~\eqref{eq:discriminant} considers these
with opposite relative signs.
This validates the observation from figure~\ref{fig:2d_masses} that the distribution of the
negative-interference events in the phase space is quite spread out and thus difficult to
separate from the horizontal band of the FCNC contribution in the $m_{W{,}\text{reco}}$
vs.\ $m_{Z{,}\text{reco}}$ plane as well as from the similarly distributed
positive-interference contribution.

\subsection{Prospects for future experiments\label{sec:results-FCC}}
We explore the potential of the interference-based approach based on various future scenarios.
These include developments in the realms of analysis methods, detector development, and future colliders.\\

\textbf{Improved $b$-tagging.}
The performance of $b$-tagging algorithms is crucial for the suppression of background contributions.
This is evident when considering that the main background contribution after the event selection (see  section \ref{sec:samples})
is $t\overline{t} \to b\overline{s}c\,\mu^-\nu_\mu\overline{b}$, which only differs from the signal final state
by an $\overline{s}$ instead of a $\overline{b}$ quark.
Thus, we expect a gain in sensitivity with increasing light-jet rejection factors at the considered $b$-tagging working points.
The $b$-tagging algorithms that provide this rejection are being constantly improved by the experimental collaborations.
An approach based on Graph Neural Networks~\cite{GNN} has already shown increased performance in comparison to traditional approaches.
To examine the effects of improved $b$-tagging algorithms, the analysis is repeated with light-jet rejection rates multiplied by a factor of two.
The resulting exclusion limit is
\begin{equation}
  g_{\text{excl}}^{\text{tag}} = \tol{8.0}{1.6}{1.2}\times 10^{-3}.
\end{equation}
This amounts to a relative improvement of the expected limit of around 9$\%$ compared to
the baseline result presented in section~\ref{sec:results-HL-LHC}.

\textbf{Improved jet-energy resolution.}
As discussed in section~\ref{sec:interference}, the reconstruction of the Dalitz variables
$m_{c\bar b}^2$ and $m_{b\bar b}^2$ enables the separation of the different contributions to
the parton-level $t \to b \overline{b} c$ decay.
However, for the full process, $t\overline{t} \to bq\overline{q}'\,\mu^-\nu_\mu\overline{b}$, the separation power degrades due to the
choice of wrong jet combinations in the reconstruction of the invariant masses and the limited jet-energy resolution.
Significant improvements in the resolution are expected for experiments at the FCC-hh~\cite{FCC:2018vvp}
based on simulation studies for calorimetry \cite{Aleksa:2019pvl}.
To investigate the impact of this improvement, we scale the expected limit for a jet
$p_{\mathrm{T}}$ resolution by a factor of \textonehalf~without changing any other parameter.
This results in
\begin{equation}
  g_{\text{excl}}^{\text{res}} = \tol{7.4}{1.4}{1.2} \times 10^{-3},
\end{equation}
which  corresponds to an improvement of about 16$\%$.

\textbf{Improved statistical power.}
The FCC-hh is projected to deliver an integrated luminosity of the order of $20\,\mathrm{ab}^{-1}$ at a
centre-of-mass energy of 100~TeV~\cite{FCC:2018vvp}.
This presents an excellent opportunity to search for $tZc$ FCNC effects in the realm of small
coupling constants with the interference-based approach.
We do not generate new MC samples for $\sqrt{s} = 100\,$TeV.
Instead, we scale the event weights by a common factor of
$\sigma_{t\overline{t}}(100\,\mathrm{TeV})/\sigma_{t\overline{t}}(14\,\mathrm{TeV}) \approx 35$,
which is the increase of the $t\overline{t}$ cross section due to the higher centre-of-mass-energy
\cite{Mangano:2016jyj}, as the signal and the main background processes rely on $t\bar{t}$ production.
However, we neglect any difference in the $\sqrt{s}$ scaling of the cross sections in the presence of
additional jets for the background processes.
The projected exclusion limit for this scenario is hence a rough estimate.
Including these changes and repeating the analysis yields a limit of
\begin{equation}
	g_{\text{excl}}^{\text{stat}} = \tol{1.9}{0.5}{0.4} \times 10^{-3}\,,
\end{equation}
which amounts to an improvement of around a factor of four.

\textbf{Combination of improvements.}
As a last scenario, we  combine all three improvements discussed above.
Therefore, this scenario corresponds to a rough projection of the sensitivity at a future
general-purpose detector at the FCC-hh with significantly improved $b$-tagging algorithms and jet resolution.
Retraining and evaluating the DNN on the adjusted dataset, we obtain an expected limit of
\begin{equation}
	g_{\text{excl}}^{\text{comb}} = \tol{1.2}{0.4}{0.3} \times 10^{-3}.
\end{equation}
This corresponds to an improvement of about a factor of seven and results in an upper limit
on the branching fraction of $\mathcal{B}_{\mathrm{excl}}^{\text{comb}}(t \rightarrow Zc) = 1.2 \times 10^{-6}$.

\subsection{Comparison to other approaches\label{sec:other_approaches}}

We compare the sensitivity of the interference-based approach to other approaches that target $tZc$ FCNC effects.
We briefly introduce three alternative approaches and then discuss the relative sensitivities of the different methods.\\

\textbf{Leptonic analysis.}
Traditionally, $tZq$ FCNCs are searched for by using the leptonic $Z \to \ell^+ \ell^-$ decay mode instead
of the hadronic decay $Z \to b\overline{b}$. This leads to three-lepton final states for the signal,
which are associated with low SM-background contributions.
Ref.~\cite{ATLAS:2023qzr} provides the tightest expected exclusion limit for
$\mathcal{B} (t \rightarrow Zc)$ of $11 \times 10^{-5}$ to date. It considers both single-top quark
production via an FCNC $tZc$ vertex ($q g \to t Z $ \footnote{%
We implicitly include charge-conjugated processes in the following discussions.})
and top-quark pair production with an FCNC decay of one of the top quarks.
Using the simple scaling introduced in section \ref{sec:introduction}, we obtain an expected
exclusion limit for $3000\,\mathrm{fb}^{-1}$ of
\begin{equation}
  \mathcal{B}_{\mathrm{excl}}^{\mathrm{lep}}(t \rightarrow Zc) \approx 11 \times 10^{-5} \cdot \sqrt{\frac{139}{3000}} \approx 2.4 \times 10^{-5}\,.
\end{equation}
Here, we have taken the limit for a left-handed coupling, just as in our studies, and have assumed that systematic uncertainties
will reduce according to the same scaling as the statistical uncertainties with the increase in integrated luminosity.
This simple projection shows some tension with the extrapolation in Ref.~\cite{ATL-PHYS-PUB-2019-001} of the search for $tZc$
FCNC effects with $36.1\,\mathrm{fb}^{-1}$ at $\sqrt{s}=13\,$TeV \cite{ATLAS:2018zsq} by the ATLAS collaboration,
which gives an expected upper limit of $4$ to $5 \times 10^{-5}$ for the HL-LHC, depending on the assumptions
on the reduction of systematic uncertainties. This limit is looser than the one obtained from the scaling above.
This hints at the importance of the correct estimation of the long-term reduction of systematic uncertainties and
highlights that the assumption that systematic uncertainties decrease according to the same scaling as statistical
uncertainties may indeed be over-optimistic for the leptonic approach.
The extrapolation to the FCC-hh scenario results in an expected limit of $1.6 \times 10^{-6}$, where we again have used an integrated
luminosity of $20\,\mathrm{ab}^{-1}$ and included a factor of $35$ for the increase of the cross sections
with $\sqrt{s}$, based again on the scaling of the $t\bar{t}$ cross section.
This projection is probably optimistic and we regard it as a rough estimate. In particular, the factor of
$35$ is unlikely to capture the increase of the cross section of the FCNC production mode accurately.
Additionally, this scaling implies a reduction of systematic uncertainties by a factor of more than
$15$, which does not seem realistic given the challenging experimental conditions at the FCC-hh.

\textbf{Ultraboosted approach.}
In Ref.~\cite{Aguilar-Saavedra:2017vka}, it was proposed to search for top-FCNC effects in $t\gamma$ and $tZ$ production
in the ultraboosted regime in which the decay products of the top quark merge into a single jet.
In contrast to our approach, this method is only sensitive to the production mode.
The ultraboosted approach is projected to yield an exclusion limit of $\mathcal{B}(t \rightarrow Zc) < 1.6 \times 10^{-3}$
at the HL-LHC,\footnote{We quote the significantly more sensitive semileptonic decay channel
of the top quark and do not attempt to provide a combination with the hadronic decay channel.}
considering a single source of systematic uncertainty on the number of background events of 20$\%$ \cite{Aguilar-Saavedra:2017vka}.
The projected limit for the FCC-hh is $3.5 \times 10^{-5}$ \cite{Aguilar-Saavedra:2017vka}.\footnote{
We scale the limit from
Ref.~\cite{Aguilar-Saavedra:2017vka} by $1/\sqrt{2}$ since we assume an integrated luminosity of
$20\,\mathrm{ab}^{-1}$ for the FCC-hh instead of $10\,\mathrm{ab}^{-1}$.
We perform the same rescaling for the FCC-hh projection of the triple-top-quark method.}

\textbf{Triple-top-quark production.}
Another way to search for top-quark FCNC effects is in triple-top-quark production: $q g \to t B^{*}$ with
$B^{*} \to t \bar t$ \cite{Barger:2010uw, Chen:2014ewl,Malekhosseini:2018fgp,Khanpour:2019qnw}.
In this process, a single top quark is produced alongside an off-shell boson $B^{*}$ mediating the FCNC, which splits into a $t \bar t$ pair.
The studies are performed for the same-sign lepton topology $\nu_{\ell} \ell^+ b \, q \bar{q}' \bar b \, \nu_{\ell'} \ell'^+ b$,
which benefits from the fact that SM background contributions are small.
However, as is also the case for ultraboosted $tZ$ production, the expected limit on $\mathcal{B}(t \rightarrow Zc)$ of
$1.35 \times 10^{-2}$ at the HL-LHC \cite{Malekhosseini:2018fgp} is relatively weak and has already been surpassed by
analyses from the ATLAS \cite{ATLAS:2023qzr} and CMS collaborations \cite{CMS:2017wcz} using the leptonic analysis.
The limit achievable at the FCC-hh is estimated to be $4.6 \times 10^{-4}$ \cite{Khanpour:2019qnw}.

\begin{table}[ht]
  \begin{center}
    \caption{
      Expected 95\% CL limits for the HL-LHC and FCC-hh scenarios for the presented interference-based approach,
      the approach with leptonic $Z \to \ell^+ \ell^-$ decay (scaled based on~\cite{ATLAS:2023qzr}),
      the ultraboosted approach~\cite{Aguilar-Saavedra:2017vka},
      and triple-top-quark production in the same-sign lepton channel~\cite{Malekhosseini:2018fgp,Khanpour:2019qnw}.
      The limits for the ultraboosted and the triple-top approaches from the references are scaled by $1/\sqrt{2}$
      to account for our assumption that roughly $20\,\mathrm{ab}^{-1}$ will be available at the FCC-hh.\\[-0.5em]
      \label{tab:limit_summary}}
        \begin{tabular}{lcc}
        Approach & HL-LHC ($3\,\mathrm{ab}^{-1}$) & FCC-hh ($20\,\mathrm{ab}^{-1}$)\\
        \hline\\[-1em]
        Interference  & $6.4 \times 10^{-5}$ & $1.2 \times 10^{-6}$\\
        Leptonic      & $2.4 \times 10^{-5}$ & $1.6 \times 10^{-6}$\\
        Ultraboosted  & $1.6 \times 10^{-3}$ & $3.5 \times 10^{-5}$\\
        Triple-top    & $1.4 \times 10^{-2}$ & $4.6 \times 10^{-4}$\\
        \hline
\end{tabular}
    \end{center}
\end{table}

\textbf{Discussion.}
We summarise the expected limits of the individual approaches in table~\ref{tab:limit_summary}.
The leptonic analysis yields the most stringent limit at the HL-LHC, while both the ultraboosted and
triple-top approaches perform significantly worse than the interference-based method.
This is to be expected since these two approaches use the production mode that is suppressed by the charm-quark parton distribution function.
Our projected limit for the interference-based approach at HL-LHC of $6.4 \times 10^{-5}$ is likely to degrade when including systematic uncertainties.
However, we restricted ourselves to only one analysis region with exactly four central $b$-tagged jets.
The inclusion of more signal regions would improve the sensitivity while data-driven background estimations
from dedicated control regions could mitigate the impact of systematic uncertainties.
Additionally, the inclusion of the electron channel will improve the sensitivity.

For the FCC-hh, the relative sensitivity of the interference-based approach compared to the leptonic analysis improves when compared to the HL-LHC scenario.
This highlights the power of the interference-based approach when moving towards the realm of smaller and
smaller couplings and the analysis of larger datasets with increasing statistical power.
Nevertheless, it should be recognised that the FCC-hh would operate in a regime of very high pileup:
the average number of visible interactions per bunch crossing is projected to be $\mu \sim \mathcal{O}(1000)$ \cite{FCC:2018vvp}.
This poses notable challenges for flavour tagging and analyses that focus on jets in general.
Because of this, more thorough studies with a dedicated detector simulation would be needed to assess and
compare the sensitivity of the two approaches at the FCC-hh.
The ultraboosted approach benefits significantly more from the energy gain from
$14\,\text{TeV}$ to $100\,\text{TeV}$ as the limit is estimated to improve by a factor of approximately $46$,
while the limit from triple-top-quark production is only projected to improve by a factor of around $29$.
A clear hierarchy can be deduced: The triple top-quark approach only yields an expected limit of the order
of $10^{-4}$, while the ultraboosted approach is expected to perform better by around one order of magnitude.
The interference-based approach and the leptonic analysis are both projected to push this even further to $\mathcal{O}(10^{-6})$.

It should also be noted that the $Z\rightarrow\ell\ell$ and the interference approach have a different sensitivity to
$tZc$ and $tZu$ FCNC couplings and are hence complementary.
The $Z\rightarrow\ell\ell$ analysis that focuses on the production mode is less sensitive to the $tZc$ than to
the $tZu$ coupling due to the difference in parton distribution functions.
Nevertheless, the sensitivities to the two couplings in the production mode are expected to be more
similar at FCC-hh due to the evolution of the parton distribution functions considering higher
energy scales and the tendency for lower Bjorken $x$ compared to the LHC.
In the decay mode, the $Z\rightarrow\ell\ell$ approach has similar sensitivity to both couplings
but relies on charm-quark identification for the distinction of these couplings.
In contrast, the interference approach is almost exclusively sensitive to the $tZc$ coupling.
Thus, in case an excess over the SM prediction is observed in the future, the combination of these
approaches will allow to disentangle possible effects from these two couplings.

\section{Conclusions\label{sec:conclusions}}

Top-quark FCNCs are so highly suppressed within the SM that
any observation at the LHC or planned future hadron colliders
would constitute a clear signal of physics beyond the SM.
At hadron colliders, the traditionally most promising and most employed channel to search for 
$tZq$ FCNCs uses a trilepton signature, relying on the leptonic $Z\to\ell^+\ell^-$ decay.
Since the $t\to Z q$ decay rate is quadratically proportional to the 
FCNC coupling, i.e., $\propto g^2$, the resulting sensitivity to probe $g$ scales as 
$1/\sqrt[4]{\mathcal{L_{\mathrm{int}}}}$ with the integrated luminosity 
$\mathcal{L}_\mathrm{int}$ 
(assuming systematic uncertainties are small compared to the statistical ones).
Given the large datasets expected at the HL-LHC and planned future
hadron colliders, we investigated how to improve upon this luminosity scaling with a novel strategy.

We propose to target the hadronic, three-body decay $t\to q b  \bar b$. 
In the presence of $tZq$ FCNCs, the decay receives two interfering 
contributions: one from the FCNC ($t\to q Z(\to b\bar b)$) 
and one from the SM ($t\to b W^+(\to q\bar b)$).
Since the two contributions interfere, the three-body
rate contains a term linear in the FCNC coupling, i.e., $\propto g$. 
Therefore, for sufficiently small $g$, the sensitivity to probe $g$ scales as $1/\sqrt{\mathcal{L_{\mathrm{int}}}}$ in this channel,
thus more favourably than in the traditional multi-lepton searches.
We studied the leading parametric dependencies controlling the 
kinematics of $t\to q b  \bar b$ and identified the requirements 
on the FCNC couplings that would allow leveraging the 
interference to compete and complement traditional searches.
The interference depends on the chirality and the phase of the FCNC coupling.
It is largest for a left-handed $tZq$ coupling, while for a right-handed one it 
is suppressed by the small masses of the bottom and $q$ quark. 
We have thus focussed on the latter case of left-handed  $tZq$ couplings.
The interference is active in a small kinematical
region in which both the $Z$ and $W$ bosons are ``on-shell''. 
In this small doubly-on-shell region,
we showed that the parametric dependence on $\Gamma/M$ is the same 
for the SM and the interference contribution. Therefore, targeting
this doubly-on-shell region with a dedicated search has the potential
to provide sensitivity with an improved luminosity scaling.

Based on these findings, we studied the prospects of the proposed 
search strategy for the case of left-handed FCNC $tZc$ couplings with 
constructive interference.
We consider the production of ${t\bar{t} \to c b \bar b\,\mu^-\nu_\mu\bar{b}}$ from $tZc$ FCNCs as the signal process.
We simulated this signal and relevant background processes with MadGraph5\_aMC@NLO
and emulated the detector response by smearing the parton-level objects with
resolutions similar to those at the ATLAS and CMS experiments.
We then separated the
FCNC signal processes from the backgrounds with a deep neural network that is
parameterised in the value of the FCNC coupling $g$.  This setup accounts for
the varying FCNC-interference contribution to the total FCNC signal.  If no
signs of FCNC production were found, the resulting expected 95\% confidence-level 
upper limit with the HL-LHC dataset is ${\mathcal{B}_{\mathrm{excl}}(t
\rightarrow Zc) = 6.4 \times 10^{-5}}$.  At the FCC-hh, the expected limit is
improved by up to a factor $\sim 50$, depending on the assumed detector
performance.

While this study did only consider statistical uncertainties, the effect of
systematic uncertainties should be studied in the future.  The main backgrounds
are $t\bar t$ production with light-quark jets misidentified as $b$- or
$c$-jets and $t\bar t$ production with a $W\rightarrow cb$ decay.  As in most
$t\bar t$ measurements, uncertainties in the modelling of the $t\bar t$ process
may impact the sensitivity.  The same is true for $b$-tagging and jet-related
uncertainties.  Heavy-flavour-associated $t\bar t$ production is only a minor
background and the potentially large associated systematic uncertainties are
unlikely to significantly affect the sensitivity.  Given the promising
signal-background separation of the parameterised deep neural network,
the statistical uncertainties on the number of events in the signal-dominated
phase space may still compete with the systematic uncertainties in the background contributions. 

As the integrated luminosity increases, the advantage of the new strategy over the traditional approach generally becomes more pronounced.
At the HL-LHC, the new strategy may not outperform the traditional search based on $Z\rightarrow\ell\ell$ decays.
However, at the FCC-hh, it has the potential to be competitive with the established approach.
Nevertheless, given their complementarity, the combination of the two strategies will improve
over the traditional search alone at both the HL-LHC and the FCC-hh.
Additionally, the new interference-based approach demonstrates excellent prospects compared to several other alternative proposals for top-quark FCNC searches.

Our study focussed on the case in which SM- and NP-sources of CP violation are aligned.
It would be intriguing to relax this assumption and design dedicated 
observables, e.g., asymmetry distributions, that optimally leverage the interference 
in $t\to q b\bar b$ to probe possible CP-violating phases in top-quark FCNC processes.
In general, the interference approach will be important 
to understand the nature of the anomalous coupling in case top-quark FCNCs are observed, as it also provides information on its Lorentz structure.

Given the results of our study on the proposed interference-based approach, it will be interesting to perform an analysis using current LHC data with a consistent treatment of systematic uncertainties and to estimate the sensitivity at the HL-LHC and future hadron-collider experiments under realistic experimental conditions.

\subsubsection*{Acknowledgements}
The authors thank Fady Bishara and Nuno Castro for useful comments on the manuscript.
ES thanks LianTao Wang for multiple inspiring discussions. 
The authors acknowledge the support from the Deutschlandstipendium (LC),
the German Research Foundation (DFG) Heisenberg Programme (JE),
the Studienstiftung des deutschen Volkes (JLS),
and the partial support by the Fermi
Fellowship at the Enrico Fermi Institute and the U.S.\ Department
of Energy, Office of Science, Office of Theoretical Research in High
Energy Physics under Award No.\ DE-SC0009924 (ES).

\appendix
\section{Two-body branching fractions\label{app:BRtwobody}}
Resonant $W$- and $Z$-boson production (if top FCNCs are present) dominate
the inclusive rate for the three-body decay $t\to cb\bar b$ via the diagrams
in figure~\ref{fig:3bodydecay}.
As discussed in section~\ref{sec:interference}, these contributions are well described in the
narrow-width approximation in terms of inclusive two-body decay rates.
Here, we collect the two-body decay rates in Eq.~\eqref{eq:BRincl} that enter the decay
$t\to cb\bar b$ in the SM and when an anomalous $tZc$ coupling is present:
\begin{align}
  \BR(t\to Zc)_{\text{FCNC}} &=
\frac{g^2 }{128 \pi}\frac{m_t}{\Gamma_t}\frac{m_t^2}{M_Z^2}
\left(1-\frac{M_Z^2}{m_t^2}\right)^2 \left(1+2\frac{M_Z^2}{m_t^2}\right)\,,
\\
\BR(t\to W^+ b)_{\text{SM}} &=
\frac{e^2 |V_{tb}|^2}{ 64 \pi s_w^2}
\frac{m_t}{\Gamma _t}
\frac{m_t^2}{M_W^2}
\left(1-\frac{M_W^2}{m_t^2}\right)^2 \left(1+2\frac{M_W^2}{m_t^2}\right)\,,
\\
\BR(W^+\to c \bar b)_{\text{SM}} &=
n_c\frac{e^2 \left| V_{cb}\right|^2}{48 \pi  s_w^2}\frac{M_W}{\Gamma _W}\,,
\\
\BR(Z\to b\bar b)_{\text{SM}} &=n_c
\frac{e^2}{864 \pi  c_w^2 s_w^2}\left(9-12 s_w^2+8 s_w^4\right)\frac{M_Z}{\Gamma_Z}\,,
\end{align}
with $s_w$ and $c_w$ the sine and cosine of the weak mixing angle, and $n_c=3$ the number
of colours.

\phantomsection
\addcontentsline{toc}{section}{References}
\bibliographystyle{JHEP_jls}
\bibliography{paper_tZc_onshell}

\end{document}